\documentclass[noshowpacs,amsmath,
twocolumn,
superscriptaddress,
8pt%,aps,prb
]{revtex4-1}
\usepackage[utf8]{inputenc}
\usepackage{setspace}
\usepackage{amsmath}
\usepackage{float}
\usepackage{bm}
\usepackage{ulem}
\usepackage{graphicx}
\usepackage[nearskip,margin = 0pt]{subfig}

\usepackage{verbatim}
\usepackage{amsfonts}
\usepackage{braket}
\usepackage{siunitx}
\usepackage{amssymb}
\usepackage{upgreek}
\usepackage[colorlinks,linkcolor=blue,anchorcolor=blue,citecolor=blue,urlcolor=black]{hyperref}
\usepackage{epstopdf}
\usepackage{xcolor}
\usepackage{booktabs}
\usepackage{tabularx}
\usepackage{xtab}
\usepackage{changepage}
\usepackage{ragged2e}

\DeclareGraphicsExtensions{.pdf,.eps,.png,.jpg,.mps} 

\begin{document}

\title{Excitons probe intrinsic flat band Mottness in a van der Waals heterostructure}
\author{Xinyue Huang$^{1,2,\dagger}$, Xintong Tan$^{1,\dagger}$, Haowei Chen$^{3,\dagger}$, Yingzhou Huang$^{1}$, Yushen Zhou$^{1}$, Yuchen Gao$^{1}$, Zhijie Ma$^{4}$, Chengxin Xiao$^{3,5,6}$, Kenji Watanabe$^{7}$, Takashi Taniguchi$^{8}$, Jianpeng Liu$^{9,10}$, Zuxin Chen$^{11}$, Youguo Shi$^{4,12}$, Wang Yao$^{3,13,\star}$ and Yu Ye$^{1,9,14\star}$\\
\vspace{6pt}
$^1$State Key Laboratory for Mesoscopic Physics and Frontiers Science Center for Nano-optoelectronics, School of Physics, Peking University, Beijing 100871, China\\
$^2$Academy for Advanced Interdisciplinary Studies, Peking University, Beijing 100871, China\\
$^3$New Cornerstone Science Laboratory, Department of Physics, The University of Hong Kong, Hong Kong, China\\
$^4$University of Chinese Academy of Sciences, Beijing 100049, China\\
$^5$HK Institute of Quantum Science \& Technology, The University of Hong Kong, Hong Kong, China\\
$^6$School of Electrical and Information Engineering, Zhengzhou University, Zhengzhou, Henan 450001, China\\
$^7$Research Center for Electronic and Optical Materials, National Institute for Materials Science, 1-1 Namiki, Tsukuba 305-0044, Japan\\
$^8$Research Center for Materials Nanoarchitectonics, National Institute for Materials Science, 1-1 Namiki, Tsukuba 305-0044, Japan\\
$^9$Liaoning Academy of Materials, Shenyang 110167, China\\
$^{10}$School of Physical Science and Technology, ShanghaiTech University, Shanghai 201210, China\\
$^{11}$School of Semiconductor Science and Technology, South China Normal University, Foshan 528225, China\\
$^{12}$Songshan Lake Materials Laboratory, Dongguan 523808, China\\
$^{13}$State Key Laboratory of Optical Quantum Materials, The University of Hong Kong, Pokfulam Road, Hong Kong SAR, China\\
$^{14}$Yangtze Delta Institute of Optoelectronics, Peking University, Nantong 226010, China\\
\vspace{3pt}
$^{\dagger}$These authors contributed equally to this work.\\
$^{\star}$Corresponding to: wangyao@hku.hk; ye\_yu@pku.edu.cn}

\begin{abstract}
\begin{adjustwidth}{-2cm}{0cm}
\textbf{ABSTRACT:}
 Excitons provide a sensitive optical probe of electronic correlations in nearby two-dimensional materials, yet their coupling to intrinsic flat-band Mott systems remains largely unexplored. Here we combine gate-tunable optical spectroscopy with first-principles calculations to study monolayer WSe$_2$ in direct contact with the van der Waals Mott insulator Nb$_3$Cl$_8$. The gate evolution of WSe$_2$ excitonic resonances reveals signatures of a correlation-reconstructed Mott gap in Nb$_3$Cl$_8$ that is absent from the single-particle band picture. In the electron-doped regime, the WSe$_2$ 2s Rydberg exciton undergoes a multistage evolution and develops into interlayer attractive and repulsive polaron branches, showing that a Rydberg exciton can be dressed by strongly correlated flat-band electrons in an adjacent Mott layer. Under an out-of-plane magnetic field, spin-polarized Nb$_3$Cl$_8$ states further induce valley-selective exciton coupling, producing a strongly enhanced circular polarization of the WSe$_2$ exciton emission. These results extend exciton-based sensing and exciton-polaron physics to intrinsic flat-band Mott materials, providing an optical route to probe and engineer correlation-driven interfacial quasiparticles.

\end{adjustwidth}
\end{abstract}
\date{\today}
\maketitle

\noindent
Flat electronic bands provide a route to quantum matter in which quenched kinetic energy allows Coulomb interactions to dominate, giving rise to Mott insulating states, Wigner crystals, unconventional superconductivity, and other correlation-driven phases at the forefront of condensed-matter physics\cite{Mott1937Discussion,Wigner1934Interaction,Cao2018Unconventional,checkelsky2024flat}. In van der Waals (vdW) materials, flat bands can be engineered by moir\'e superlattices, where twist angle and lattice mismatch create narrow minibands with tunable filling and bandwidth\cite{bistritzer2011moire,wu2018hubbard}. These artificial flat-band systems have transformed the study of correlated electrons, yet their emergent phases are often inferred indirectly from transport or capacitance. A more direct optical pathway to flat-band correlations would open a complementary means of probing and controlling interaction-driven quantum phases at the atomic scale.

Excitons in atomically thin semiconductors offer such an optical pathway\cite{wang2018colloquium,mueller2018exciton,wilson_excitons_2021}. Owing to the reduced dielectric screening and the nanometer-scale spatial extent of their electron-hole Coulomb interaction, two-dimensional (2D) excitons are highly sensitive to nearby dielectric screening, charge compressibility, and interlayer Coulomb interactions\cite{chernikov2014exciton}. This sensitivity becomes especially pronounced for excited Rydberg excitons, whose enlarged wave functions allow them to act as remote optical sensors of electronic states in adjacent layers\cite{regan2022emerging}. Recent experiments have used Rydberg excitons to detect correlated insulating states, Wigner-Mott transitions, and moir\'e miniband physics in neighboring moir\'e systems, establishing exciton spectroscopy as a non-invasive probe of interfacial many-body electronic responses\cite{xu2020correlated,xu_creation_2021,hu2023observation}. In parallel, excitons coupled to a Fermi Sea can form attractive and repulsive exciton polarons\cite{Efimkin2017TrionAbsorption,Sidler2017Fermi,Liu2021ExcitonPolaron}, revealing how optical excitations are dressed by doped carriers in 2D systems.

\begin{figure*}[tbh]    
\includegraphics[width=2\columnwidth]{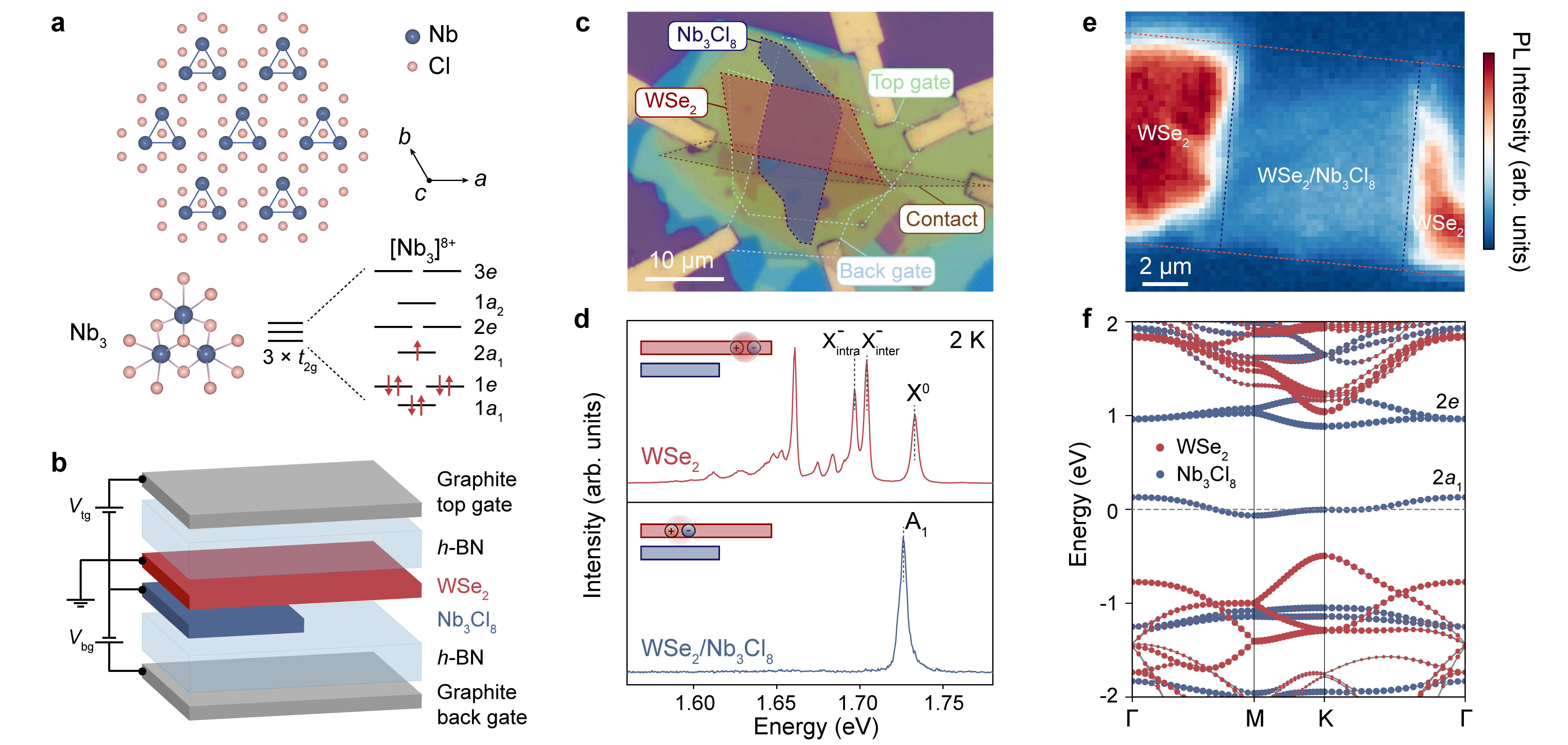}
\captionsetup{singlelinecheck=off, justification = RaggedRight}
\caption{\label{Figure1}\textbf{WSe$_2$ excitons coupled to an intrinsic flat band Mott insulator.}
     \textbf{(a)} Upper panel, top view of the Nb$_3$Cl$_8$ crystal structure, with Nb$_3$ trimer units highlighted by triangles. Lower panel, molecular orbital diagram of an isolated Nb$_3$Cl$_{13}$ cluster, showing the seven valence electrons of one Nb$_3$ trimer.
     \textbf{(b)} Schematic of the dual-gated WSe$_2$/Nb$_3$Cl$_8$ device.
     \textbf{(c)} Optical micrograph of device 1. The WSe$_2$ monolayer and Nb$_3$Cl$_8$ flake are outlined in red and blue, respectively.
     \textbf{(d)} Low-temperature PL spectra measured at 2 K from WSe$_2$-only region and the WSe$_2$/Nb$_3$Cl$_8$ heterostructure. X$^0$ denotes the neutral exciton, $\mathrm{X}_{\mathrm{inter}}^-$ and $\mathrm{X}_{\mathrm{intra}}^-$ denote the negatively charged intervalley and intravalley trions, and the single emission feature in the heterostructure is labeled as A$_1$. 
     \textbf{(e)} Spatial PL intensity map integrated over the energy range from 1.72 eV to 1.74 eV.
     \textbf{(f)} DFT-calculated band structure of the WSe$_2$/Nb$_3$Cl$_8$ heterostructure without electronic correlation effects. 
}
\end{figure*}

Despite this progress, excitonic access to correlated flat-band physics has so far been dominated by externally engineered electronic platforms, including moir\'e bands and Landau-quantized electron gases\cite{xu_creation_2021,popert2022optical,cui2024interlayer}. By contrast, intrinsic flat-band Mott materials remain largely unexplored. In these systems, electron localization arises inherently from the native lattice geometry and orbital structure of the crystal, rather than being induced by an imposed moir\'e potential or magnetic field\cite{lieb1989two,cualuguaru2022general,checkelsky2024flat,neves2024crystal}. Strong correlations can split a half-filled flat band into lower and upper Hubbard bands, producing localized electrons that differ qualitatively from the weakly interacting doped carriers usually considered in exciton-polaron physics. Whether excitons in a neighboring semiconductor can optically sense this Hubbard-band electronic structure, and whether a Rydberg exciton can be dressed by such intrinsic flat-band electrons to form interlayer many-body quasiparticles, remains unknown.

Here we address these questions by coupling monolayer WSe$_2$ to Nb$_3$Cl$_8$, a vdW breathing-kagome compound. This material is identified as an intrinsic single-band Mott insulator, driven by a half-filled flat band derived from the Nb trimer orbitals (Figure \ref{Figure1}a)\cite{gao2023discovery,liu2025direct,yang2025evidence,grytsiuk_nb3cl8_2024}. Using gate-tunable optical spectroscopy together with first-principles calculations, we show that WSe$_2$ excitons provide an optical readout of the correlation-reconstructed band alignment of the WSe$_2$/Nb$_3$Cl$_8$ heterostructure. The gate evolution of the excitonic resonances reveals the emergence of a Mott gap in Nb$_3$Cl$_8$ that is absent in the single-particle picture. In the electron-doped regime, the WSe$_2$ 2s Rydberg exciton splits into attractive and repulsive interlayer polaron branches, suggesting that an excited exciton can be dressed by strongly correlated flat-band electrons in the adjacent Mott layer. Under an out-of-plane magnetic field, we further observe valley-selective exciton coupling, manifested by a strong enhancement of circular polarization in photoluminescence. These results establish intrinsic flat-band Mott insulators as active building blocks for exciton-correlation physics and demonstrate Rydberg excitons as sensitive optical probes of Mottness in vdW heterostructures.

\begin{figure*}[tbh]    
\includegraphics[width=2\columnwidth]{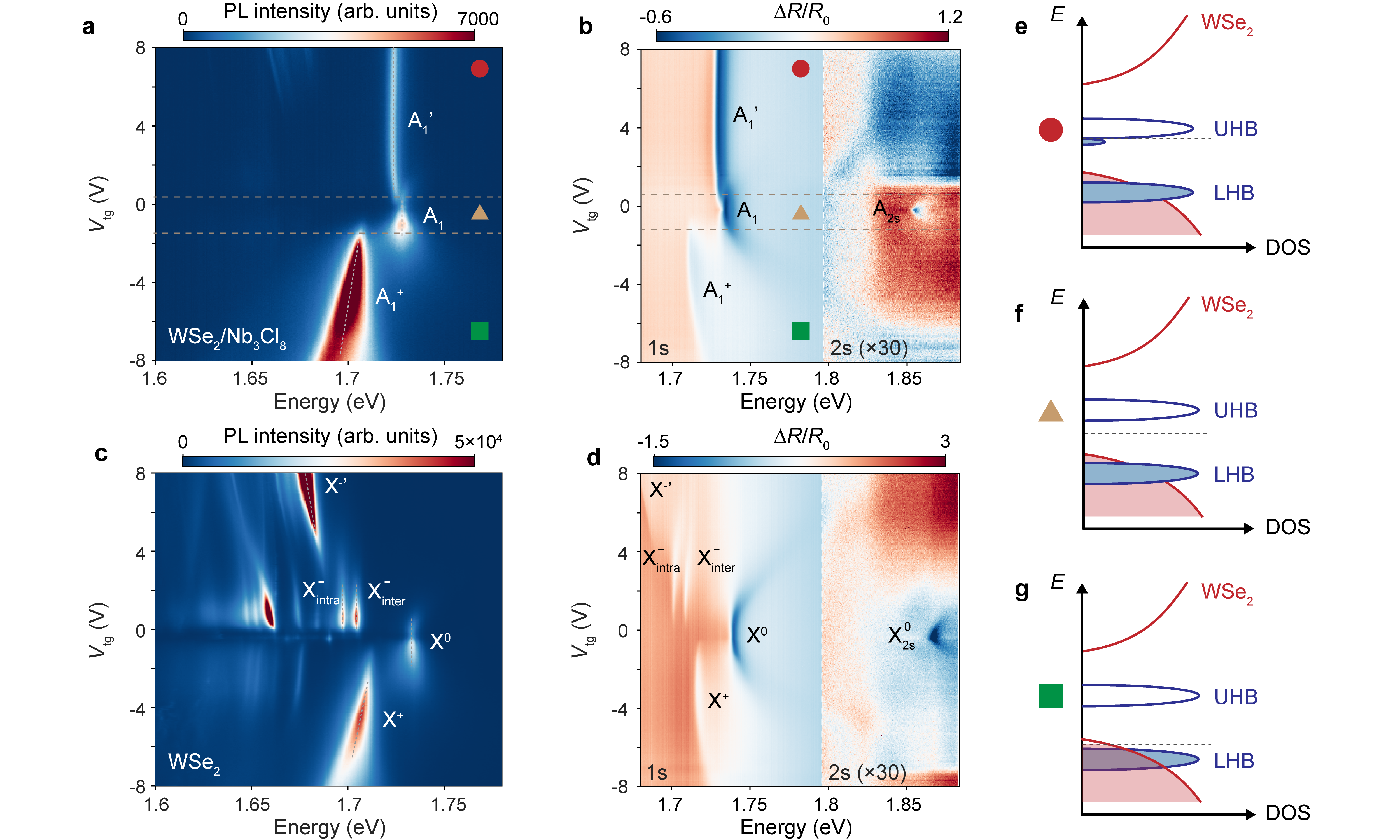}
\captionsetup{singlelinecheck=off, justification = RaggedRight}
\caption{\label{Figure2}\textbf{Gate-tunable excitonic signatures of a correlation-reconstructed Mott gap.}
     \textbf{(a,b)} PL spectra (a) and reflection contrast spectra $\Delta\textit{R}/\textit{R}_0$ (b) of the WSe$_2$/Nb$_3$Cl$_8$ heterostructure as a function of the top-gate voltage $\textit{V}_{\text{tg}}$. 
     \textbf{(c,d)} Corresponding PL spectra (c) and reflection contrast spectra $\Delta\textit{R}/\textit{R}_0$ (d) from the WSe$_2$-only region of the same monolayer. The labeled excitonic resonances are discussed in the main text. $\mathrm{X}^{-\prime}$ denotes the six-particle many-body exciton state, or hexciton\cite{li_many-body_2021,van_tuan_six-body_2022,choi_emergence_2024}. All spectra were measured at 2 K.
     \textbf{(e-g)} Schematic band alignments between WSe$_2$ and Nb$_3$Cl$_8$ in the electron-doped regime(e), near charge neutrality (f), and hole-doped regime (g). LHB and UHB denote the lower and upper Hubbard bands of Nb$_3$Cl$_8$, respectively. 
}
\end{figure*}

\bigskip
\noindent \textbf{Excitonic signatures of a correlation-reconstructed Mott gap.} To optically access the excitonic response to an intrinsic flat-band Mott layer, we fabricated dual-gated WSe$_2$/Nb$_3$Cl$_8$ heterostructures encapsulated by hexagonal boron nitride (\textit{h}-BN) and controlled by few-layer graphite (FLG) gates (Figure \ref{Figure1}b). In the main device, monolayer WSe$_2$ is partially contacted by an 18.7-nm-thick Nb$_3$Cl$_8$ flake, leaving a region of the same WSe$_2$ monolayer outside the Nb$_3$Cl$_8$ contact area (Figure \ref{Figure1}c and Supplementary Information Figure S1). This geometry enables a direct side-by-side comparison between the WSe$_2$-only region and the WSe$_2$/Nb$_3$Cl$_8$ heterostructure region within the same device, allowing the excitonic modifications induced by direct contact with Nb$_3$Cl$_8$ to be isolated from sample-to-sample variations. Consistent behavior is observed in an additional device, as shown in Supplementary Information Figures S2-S6.

Photoluminescence (PL) spectra reveal distinct emission characteristics between WSe$_2$-only region and the WSe$_2$/Nb$_3$Cl$_8$ heterostructure region (Figure \ref{Figure1}d). The measurements were performed at 2 K with grounded FLG gates and an excitation power of 65 $\mu$W. The WSe$_2$-only region exhibits multiple emission features, with the peak near 1.733 eV attributed to the neutral exciton $\text{X}^0$. The presence of negatively charged inter- and intravalley trions, $\mathrm{X}_{\mathrm{inter}}^-$ and $\mathrm{X}_{\mathrm{intra}}^-$, indicates residual electron doping in the as-fabricated monolayer\cite{PhysRevB.96.085302,barbone2018charge,PhysRevLett.123.027401}. In contrast, the heterostructure region displays a single emission line centered at approximately 1.726 eV, which we assign to the $\text{A}_1$ emission. The absence of low-energy trion features suggests that WSe$_2$ becomes nearly charge neutral upon direct contact with Nb$_3$Cl$_8$. Moreover, pronounced PL quenching is observed in the heterostructure region, as confirmed by spatial PL map in Figure \ref{Figure1}e, consistent with efficient interfacial charge transfer and/or non-radiative recombination channels at the interface\cite{hong_ultrafast_2014,rivera_observation_2015}.

Density functional theory (DFT) calculations that neglect electronic correlations place the half-filled Nb$_3$Cl$_8$ flat band, derived from the 2$a_1$ molecular orbital, inside the WSe$_2$ band gap (Figure \ref{Figure1}f). In this single-particle picture, the large density of states (DOS) of the 2$a_1$ flat band would rigidly pin the Fermi level, because complete filling or depletion of this band would require a carrier density on the order of 10$^{14}$ cm$^{-2}$\cite{gao_designer_2025}, far beyond the range accessible by electrostatic gating in our devices. The optical response of WSe$_2$ would therefore be expected to remain nearly gate independent. By contrast, gate-dependent PL and reflection contrast spectra reveal three distinct and tunable excitonic features in the WSe$_2$/Nb$_3$Cl$_8$ heterostructure (Figures \ref{Figure2}a,b). This discrepancy demonstrates that a single-particle band alignment is insufficient and that electronic correlations in Nb$_3$Cl$_8$ must be included.

\begin{figure*}[tbh]    
\includegraphics[width=2\columnwidth]{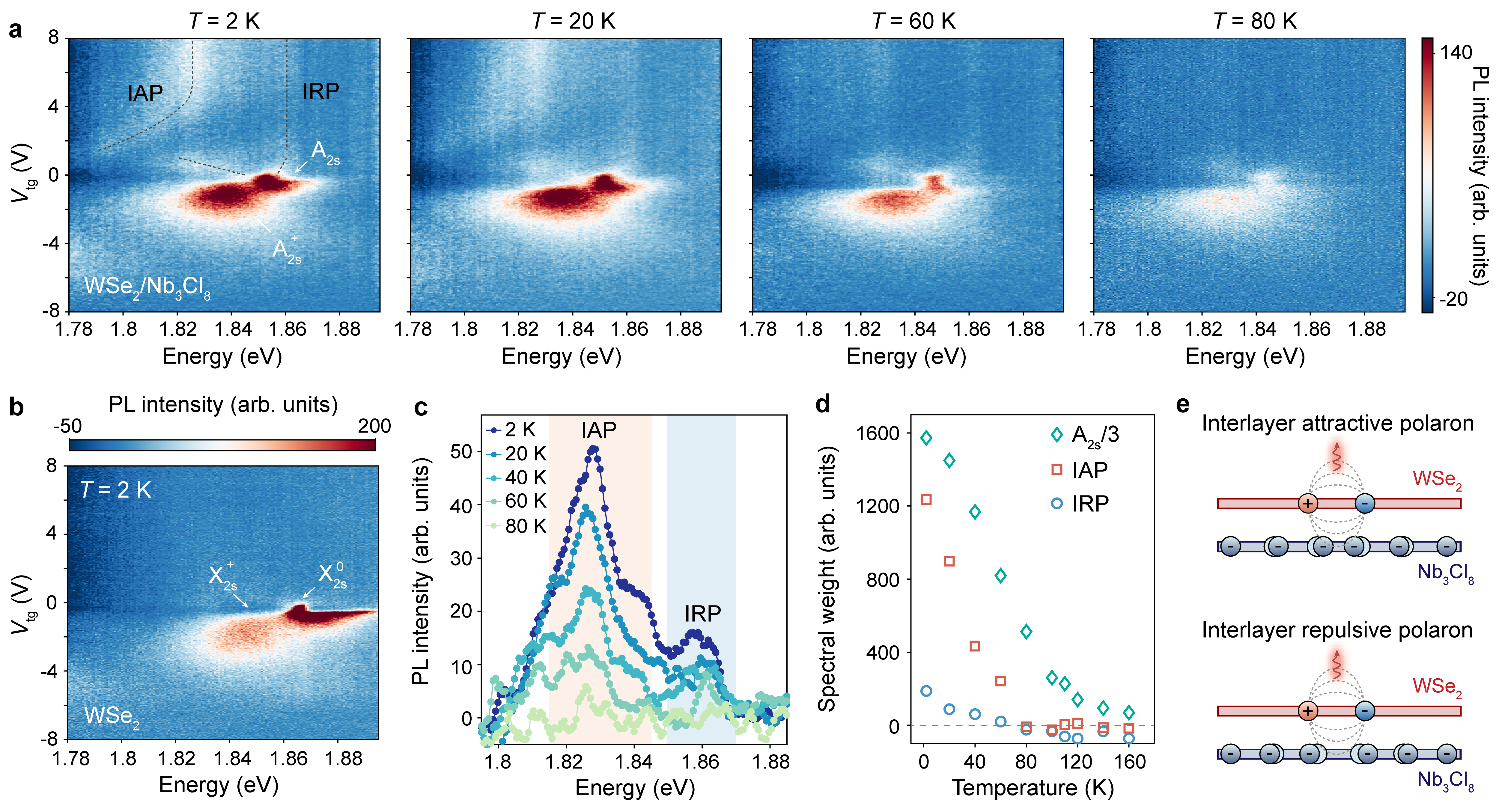}
\captionsetup{singlelinecheck=off, justification = RaggedRight}
\caption{\label{Figure3}\textbf{Rydberg interlayer attractive and repulsive polarons.}  
\textbf{(a)} Gate-dependent PL spectra in the WSe$_2$ 2s exciton energy range of the WSe$_2$/Nb$_3$Cl$_8$ heterostructure, measured at 2 K, 20 K, 60 K, and 80 K. Grey dashed guide lines mark the development of interlayer attractive and repulsive polaron branches, labeled IAP and IRP.  
\textbf{(b)} Gate-dependent PL spectra of the WSe$_2$-only region at 2 K for comparison.  
\textbf{(c)} PL line cuts averaged over $V_{\text{tg}}$ from 6 V to 8 V at different temperatures. Orange and blue shaded regions indicate the energy windows used to extract the IAP and IRP spectral weights, respectively.  
\textbf{(d)} Temperature dependence of the spectral weights of A$_{\text{2s}}$, IAP, and IRP. The A$_{\text{2s}}$ spectral weight was obtained by integrating the background-subtracted PL intensity from 1.82 eV to 1.87 eV over $V_{\text{tg}}$ from $-$0.8 V to 0.6 V. The IAP and IRP spectral weights were obtained from the energy windows defined in \textbf{(c)} over $V_{\text{tg}}$ from 6 V to 8 V. 
\textbf{(e)} Schematic illustration of interlayer attractive and repulsive polarons formed by dressing a WSe$_2$ Rydberg exciton with doped electrons in adjacent Nb$_3$Cl$_8$.
}
\end{figure*}

In the strong-correlation regime, where the interaction strength $U$ exceeds the bandwidth $W$, the half-filled flat band of Nb$_3$Cl$_8$ is split into a filled lower Hubbard band (LHB) and an empty upper Hubbard band (UHB), separated by a Mott gap\cite{PhysRevB.44.943}. This correlation-driven reconstruction of the Nb$_3$Cl$_8$ spectrum reshapes the band alignment with WSe$_2$ and provides a natural framework for understanding the gate-dependent excitonic resonances. We therefore focus on the response under top-gate voltage $V_\text{tg}$, for which the evolution of the excitonic features defines three regimes: electron-doped, charge-neutral, and hole-doped, separated by $V_\text{tg}=0.5$ V and $-1.3$ V (Figures \ref{Figure2}e-g). The weak response to back-gate modulation is shown in Supplementary Information Figure S7.  

At charge neutrality, the heterostructure exhibits two resonances at 1.733 eV and 1.853 eV, labeled A$_1$ and A$_{\text{2s}}$, respectively (Figures \ref{Figure2}a,b). Their energies are consistent with the 1s and 2s neutral excitons of monolayer WSe$_2$ measured in the WSe$_2$-only region (Figures \ref{Figure2}c,d). In the hole-doped regime, the Mott gap places the LHB below the WSe$_2$ valence-band maximum, so that injected holes preferentially occupy the WSe$_2$ valence band (Figure \ref{Figure2}g). This hole doping gives rise to the positive attractive polaron A$_1^+$, analogous to the X$^+$ state in pristine monolayer WSe$_2$.

In the electron-doped regime, electrons are instead injected into Nb$_3$Cl$_8$. For a doped Mott insulator, such carrier injection can transfer spectral weight from high-energy Hubbard bands to emergent low-energy states near the Fermi level (Figure \ref{Figure2}e)\cite{katsufuji1995spectral,cai2016visualizing}. This correlation-driven redistribution allows the Fermi level to remain within the WSe$_2$ band gap, so that the WSe$_2$ resonance A$_1'$ retains a neutral-exciton-like character and shows only weak energy variation with further electron doping. The assignments of these excitonic features are further supported by power-dependent measurements and valley-splitting analyses (Supplementary Information Figures S8 and S9). Notably, the onset of UHB occupation is accompanied by an immediate redshift of A$_1'$ by approximately 4 meV relative to A$_1$, consistent with a modified dielectric screening response\cite{gao_dynamical_2016,raja_coulomb_2017,ben_mhenni_breakdown_2025} induced by electron doping in Nb$_3$Cl$_8$.

\bigskip
\noindent \textbf{Rydberg interlayer polarons driven by correlated flat-band electrons.} Given the enhanced sensitivity of the spatially extended Rydberg excitons to their surrounding dielectric and charge environment\cite{xu2020correlated,xu_creation_2021}, we next examine the gate evolution of the WSe$_2$ 2s exciton in the WSe$_2$/Nb$_3$Cl$_8$ heterostructure. In the hole-doped regime, the 2s resonance follows the behavior of the 1s transition. The neutral A$_{\text{2s}}$ feature evolves into a broadened and redshifted A$_{\text{2s}}^+$ state, consistent with the formation of a hole-dressed excitonic state analogous to the X$^+$ response in monolayer WSe$_2$ (first panel of Figure \ref{Figure3}a). A qualitatively different behavior appears upon electron doping, where electrons are injected into the Nb$_3$Cl$_8$ UHB while the Fermi level remains inside the WSe$_2$ band gap. During the gate-voltage interval in which A$_1$ evolves toward A$_1'$, A$_{\text{2s}}$ first exhibits a modest redshift and then separates into two emerging 2s-derived branches that shift towards opposite directions (Figure \ref{Figure3}a). The low-energy branch undergoes a pronounced redshift, reaching a total shift of nearly 65 meV relative to the neutral A$_{\text{2s}}$ resonance, whereas the high-energy branch blueshifts. At higher electron doping, where the 1s response has evolved into the A$_1'$ resonance, these two branches develop into the interlayer attractive and repulsive polaron features, IAP and IRP. In this regime, the IAP initially blueshifts before approaching a saturated energy, while the IRP remains on the higher-energy side. This multistage evolution contrasts sharply with the weak energy variation of the A$_1'$ resonance, as further summarized by the 1s/2s energy comparison in Supplementary Information Figure S10. The large and nonmonotonic response of the 2s-derived lower branch reflects its Rydberg character, because the weakly bound and spatially extended 2s exciton is far more sensitive than the 1s exciton to the remote dielectric screening and charge susceptibility of the adjacent Nb$_3$Cl$_8$ layer. We therefore associate the initially redshifting 2s-derived feature with a precursor to IAP formation rather than a conventional 2s trion resonance.

At higher electron doping, the IAP and IRP become well resolved and approach two separated energy plateaus. These features are intrinsic to the WSe$_2$/Nb$_3$Cl$_8$ heterostructure, because the WSe$_2$-only region remains featureless in the same spectral and gate-voltage range (Figure \ref{Figure3}b). Temperature-dependent PL measurements further support their many-body origin. The spectral weights of the IAP and IRP are extracted by integrating the background-subtracted PL intensity over the energy windows indicated in Figure \ref{Figure3}c. Both branches lose spectral weight rapidly with increasing temperature and are strongly suppressed by 80 K, indicating that these interlayer polaron states are less robust than the intralayer 2s neutral exciton (Figure \ref{Figure3}d and Supplementary Information Figure S11). As illustrated in Figure \ref{Figure3}e, the IAP and IRP can be viewed as interlayer analogues of exciton polarons in monolayer TMDs, arising from coupling between a WSe$_2$ Rydberg exciton and doped correlated electronic states in the adjacent Nb$_3$Cl$_8$ layer.

This interpretation is consistent with recent studies of excited-state exciton polarons, where 2s and 3s Rydberg excitons exhibit stronger density-dependent energy shifts and oscillator-strength redistribution than their 1s counterparts\cite{xu2020correlated,christianen2025interactions}. Interlayer Fermi polarons of 2s excitons have also been observed in WSe$_2$/graphene heterostructures, where Landau quantization of carriers in the adjacent graphene layer is required for the emergence of attractive and repulsive interlayer 2s polaron branches\cite{cui2024interlayer}. In contrast, the present WSe$_2$/Nb$_3$Cl$_8$ heterostructure does not rely on magnetic-field-induced Landau quantization to quench the kinetic energy of the adjacent carriers. Instead, the intrinsically flat Nb$_3$Cl$_8$ UHB may provide a correlated electronic environment in which carrier localization and Coulomb interactions are already strong. The saturated IAP-IRP separation reaches approximately 35 meV at high electron doping, about four times larger than that reported for WSe$_2$/graphene in the Landau-quantized regime, suggesting enhanced interlayer many-body interactions associated with correlated electronic states in Nb$_3$Cl$_8$. Although a quantitative theory incorporating localized Hubbard-band electrons, remote dielectric screening, and the internal structure of the Rydberg exciton is still needed, the observed onset redshift, subsequent branch splitting, thermal dissociation, and large saturated IAP-IRP separation collectively support the formation of correlation-enhanced Rydberg interlayer polarons.

\begin{figure*}[tbh]    
\includegraphics[width=2\columnwidth]{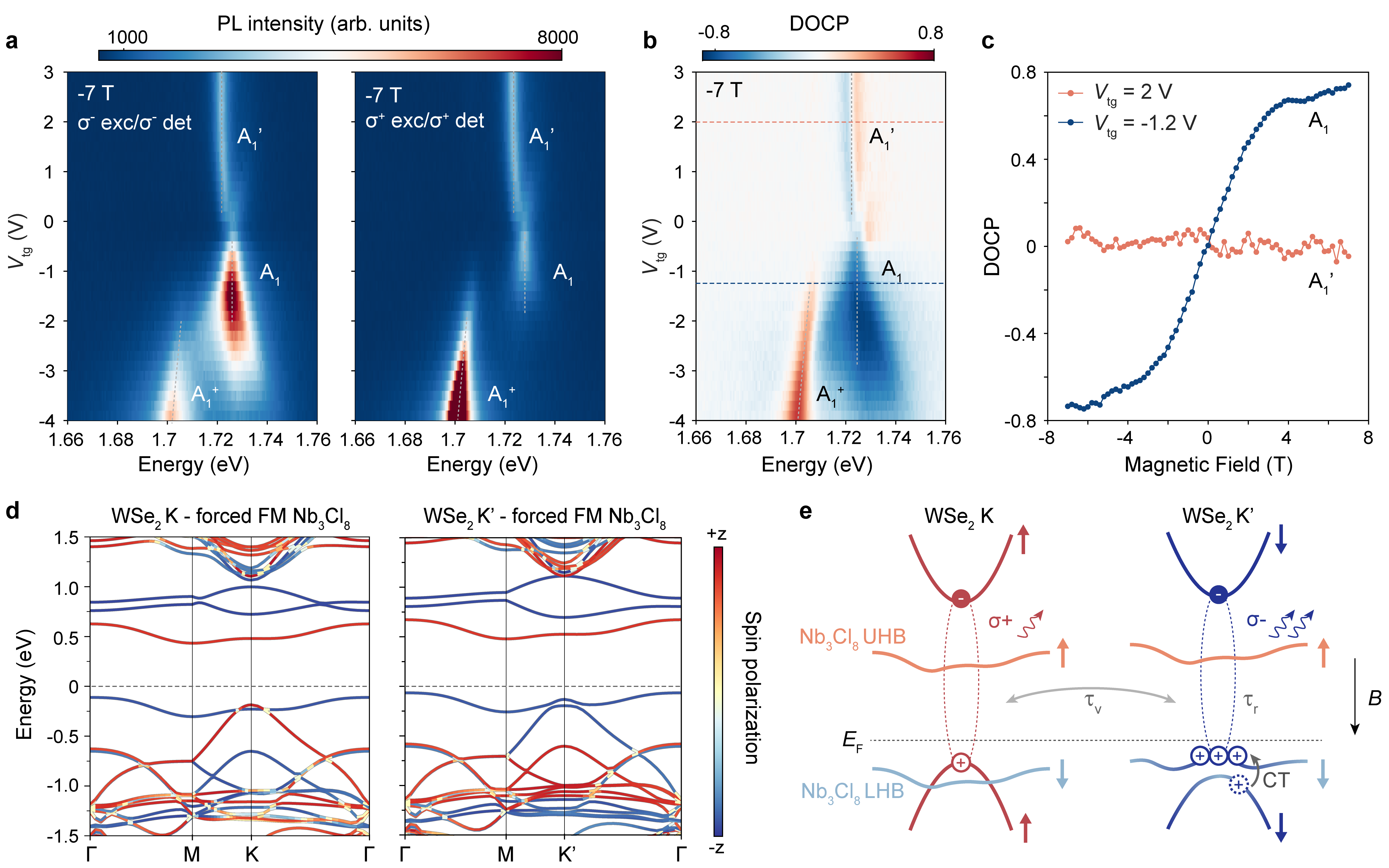}
\captionsetup{singlelinecheck=off, justification = RaggedRight}
\caption{\label{Figure4}\textbf{Valley selective exciton coupling to spin-polarized Nb$_3$Cl$_8$ states.}
     \textbf {(a)} Polarization-resolved PL intensity maps as a function of top-gate voltage $V_{\text{tg}}$ and photon energy, measured at 2 K under an out-of-plane magnetic field of $B$ = $-$7 T. The left and right panels show co-circular $\sigma^-$ excitation/detection and co-circular $\sigma^+$ excitation/detection, respectively. 
     \textbf{(b)} Degree of circular polarization map, defined as DOCP $=(I_{\sigma^+}-I_{\sigma^-})/(I_{\sigma^+}+I_{\sigma^-})$, obtained from the spectra in \textbf{(a)}.
     \textbf{(c)} Magnetic field dependence of the DOCP for A$_1'$ and A$_1$, measured at $V_{\text{tg}}$ = 2 V and $V_{\text{tg}}$ = $-$1.2 V, respectively, as marked by the orange and blue dashed lines in \textbf{(b)}. 
     \textbf{(d)} Spin-polarized band structures of the WSe$_2$/Nb$_3$Cl$_8$ heterostructure calculated using DFT+$U$ with spin-orbit coupling along momentum paths containing the WSe$_2$ K (left panel) and K$'$ (right panel) valleys.
     \textbf{(e)} Schematic of valley-dependent hybridization and exciton relaxation pathways in the heterostructure.
}
\end{figure*}

\bigskip
\noindent \textbf{Valley-selective coupling to spin-polarized flat band.} Beyond charge correlations, Nb$_3$Cl$_8$ hosts localized spin-1/2 moments on each Nb trimer, which form an effective triangular lattice with strong magnetic frustration (Figure \ref{Figure1}a) \cite{liu_possible_2024,fernando_strain-tunable_2026}. When coupled to monolayer WSe$_2$, where spin and valley degrees of freedom are locked by broken inversion symmetry and spin-orbit coupling, these localized moments provide an opportunity to examine valley-dependent proximity coupling between WSe$_2$ excitons and magnetic flat-band states of Nb$_3$Cl$_8$.

Figure \ref{Figure4}a shows gate-dependent polarization-resolved PL measured at 2 K under an out-of-plane magnetic field of $B = -7$ T, using co-circular excitation and detection configurations ($\sigma^-$/$\sigma^-$ and $\sigma^+$/$\sigma^+$). We quantify the degree of circular polarization by DOCP $=(I_{\sigma^+}-I_{\sigma^-})/(I_{\sigma^+}+I_{\sigma^-})$. The resulting DOCP map reveals strongly state-dependent valley polarization in the WSe$_2$/Nb$_3$Cl$_8$ heterostructure (Figure \ref{Figure4}b). In the electron-doped regime, the A$_1'$ resonance remains nearly unpolarized. By contrast, the A$_1$ resonance at small negative $V_{\text{tg}}$, despite sharing a common neutral exciton nature as A$_1'$, shows a pronounced circular polarization. The positively charged A$_1^+$ feature exhibits an opposite valley polarization to A$_1$, because A$_1^+$ involves intervalley pairing between the recombining electron-hole pair and the additional hole\cite{jones2013optical}. Magnetic-field-dependent measurements further highlight this contrast (Figure \ref{Figure4}c). While the DOCP of A$_1'$ remains close to zero from $-$7 T to 7 T, the DOCP of A$_1$ increases rapidly with the magnetic field and reaches approximately 0.8 at 7 T.

Comparison with the WSe$_2$-only region clarifies the role of Nb$_3$Cl$_8$. The field-dependent DOCP of A$_1'$ closely resembles that of the neutral exciton X$^0$ in monolayer WSe$_2$ near charge neutrality, remaining weak over the measured field range (see Supplementary Information Figure S12). This behavior supports the assignment of A$_1'$ as a neutral-exciton-like WSe$_2$ resonance in the electron-doped regime. In contrast, although A$_1$ at small negative $V_{\text{tg}}$ is also neutral-exciton-like in energy and corresponds to the X$^0$ resonance under moderate hole doping, its DOCP is strongly enhanced relative to the monolayer case (Supplementary Information Figure S12). This enhancement identifies the proximate Nb$_3$Cl$_8$ layer as the origin of the modified valley dynamics of A$_1$.

The microscopic origin of this enhancement is suggested by spin-polarized DFT+$U$ calculations including spin-orbit coupling. Under a large negative magnetic field, the localized moments in Nb$_3$Cl$_8$ are expected to become partially aligned, producing spin-polarized Hubbard bands. Although the Nb$_3$Cl$_8$ bands themselves do not distinguish between the K and K$'$ valleys, spin-valley locking in WSe$_2$ makes their hybridization with Nb$_3$Cl$_8$ strongly valley selective. For the spin configuration shown in Figure \ref{Figure4}d, the WSe$_2$ valence-band edge in the K valley has spin character opposite to the nearby Nb$_3$Cl$_8$ states, suppressing interlayer hybridization. In the K$'$ valley, the WSe$_2$ and Nb$_3$Cl$_8$ states share the same spin orientation, enabling stronger interlayer mixing and spin-allowed charge transfer. The resulting valley-asymmetric coupling modifies the recombination and relaxation pathways of WSe$_2$ excitons.

This picture provides a natural explanation for the enhanced DOCP. In steady-state PL, the circular polarization can be approximated as $P_c=\frac{P_0}{1+\frac{\tau}{\tau_v}}$ \cite{wang2018colloquium}, where $P_0$ is the optically initialized polarization, $\tau$ is the exciton lifetime, and $\tau_v$ is the intervalley scattering time\cite{jones2013optical,glazov_exciton_2014,yu_valley_2014}. Valley-selective hybridization with spin-polarized Nb$_3$Cl$_8$ states can reduce the exciton lifetime in the strongly coupled valley by opening additional interlayer relaxation or charge-transfer channels (K$'$ valley in Figure \ref{Figure4}e). A shorter population lifetime of the A$_1$ exciton state means the
presence of a larger valley polarization upon its emission. This mechanism is consistent with the strong A$_1$ polarization, the weak A$_1'$ polarization, and the valley-selective band hybridization revealed by calculations. Together, these results show that the magnetic correlated electronic states in Nb$_3$Cl$_8$ can selectively couple to WSe$_2$ valleys, leading to spin-dependent modifications of exciton valley dynamics in the heterostructure.

\bigskip
\noindent 
\textbf{Conclusion}\\
We have shown that excitons in monolayer WSe$_2$ provide an optical readout of Mottness in the adjacent flat-band Mott insulator Nb$_3$Cl$_8$. Gate-tunable spectroscopy reveals a correlation-reconstructed Mott gap that cannot be captured by a single-particle band alignment, while the spatially extended 2s Rydberg exciton evolves into interlayer attractive and repulsive polaron branches when electrons are doped into the Nb$_3$Cl$_8$ upper Hubbard band. Under an out-of-plane magnetic field, spin-polarized Nb$_3$Cl$_8$ states further produce valley-selective coupling to WSe$_2$ excitons, resulting in a strongly enhanced circular polarization of the A$_1$ emission. These observations establish intrinsic flat-band Mott layers as active quantum materials for exciton-based sensing, dressing, and valley control. More broadly, they open an optical route to interrogate and engineer correlation-driven quasiparticles in van der Waals heterostructures beyond moir\'e electronic platforms.

\bigskip

\bigskip
\noindent 
\textbf{Methods}\\
\noindent \textbf{Crystal Synthesis.} 
WSe$_2$ single crystals were synthesized by the chemical vapor transport method. High-purity W and Se powders (99.99\%, Aladdin) were mixed in a stoichiometric molar ratio of 1:2. Iodine (5 mg mL$^{-1}$) was used as the transport agent. The mixture was sealed under vacuum in a quartz ampoule (120 mm length, 16 mm inner diameter). Crystal growth was carried out in a two-zone furnace with source and growth zones held at 1200 $^\circ$C and 960 $^\circ$C, respectively, for 9 days. The resulting WSe$_2$ crystals were collected from the cold zone. 

Single crystals of Nb$_3$Cl$_8$ were grown using a PbCl$_2$ flux-assisted method. Nb$_3$Cl$_8$ precursor powder was first synthesized by a solid-state reaction involving high-purity Nb (Alfa Aesar, 99.99\%) and NbCl$_5$ (Alfa Aesar, 99.9\%). The Nb and NbCl$_5$ powders were mixed at a molar ratio of 7:8, sealed under vacuum in a quartz tube, and annealed at 700 $^\circ$C for 48 h. The as-synthesized Nb$_3$Cl$_8$ precursor was then mixed with excess PbCl$_2$ flux and sealed again under vacuum. The mixture was heated to 750 $^\circ$C for 20 h, held at 750 $^\circ$C for 300 h, and then cooled to 500 $^\circ$C over 100 h. After natural cooling to room temperature, Nb$_3$Cl$_8$ crystals were isolated by removing the residual flux with hot deionized water.
\\

\noindent \textbf{Device Fabrication.} 
Monolayer WSe$_2$, Nb$_3$Cl$_8$, few-layer graphite, and \textit{h}-BN flakes were first mechanically exfoliated from bulk crystals onto Si/SiO$_2$ substrates under ambient conditions. Monolayer WSe$_2$ flakes were identified by optical contrast and further confirmed by room-temperature PL spectroscopy. WSe$_2$/Nb$_3$Cl$_8$ heterostructures were assembled by a dry transfer method using a poly (bisphenol A carbonate) film mounted on a dome-shaped polydimethylsiloxane (PDMS) stamp. For electrical contacts, 5 nm Cr and 50 nm Au electrodes were patterned using electron-beam lithography, followed by reactive ion etching in a CHF$_3$/O$_2$ plasma, electron-beam evaporation, and standard lift-off procedures. \\

\noindent \textbf{Optical Measurements.} 
Optical measurements were performed in a closed-cycle helium cryostat (attoDRY2100) with a base temperature of 1.6 K, equipped with a superconducting magnet that provided an out-of-plane magnetic field up to 9 T. PL measurements were carried out in a backscattering geometry using a HeNe laser with a photon energy of 1.96 eV as the excitation source. The excitation beam was focused onto the sample through a high numerical aperture objective with $\text{NA} = 0.82$, giving a beam spot size of approximately 1 $\mu$m in diameter. The emitted light was collected by a spectrometer, SpectraPro HRS-500S, dispersed by a grating with a 150 grooves/mm grating, and detected using a liquid-nitrogen-cooled charge-coupled device (PyLoN:400). Spatial PL maps were obtained by scanning the sample with a high-precision $x-y$ piezoelectric stage. For reflection contrast measurements, broadband white light from a halogen lamp was spatially filtered through a 50 $\mu$m diameter pinhole (JCOPTIX MPS1-50) and collimated by an objective lens. The resulting beam had a spot diameter of approximately 2 $\mu$m and an excitation power below 100 nW to minimize sample heating or photoinduced effects.\\

\noindent \textbf{First-principles calculations.}
The electronic structure of the WSe$_2$/Nb$_3$Cl$_8$ heterostructure was calculated using the Vienna Ab initio Simulation Package (VASP)\cite{PhysRevB.54.11169} with projector augmented wave pseudopotentials\cite{PhysRevB.50.17953} and a plane wave basis. The heterostructure was constructed by doubling the lattice of monolayer WSe$_2$ to match the lattice constant of the monolayer Nb$_3$Cl$_8$. After structural relaxation, the in-plane lattice constant of the heterostructure was 6.63 $\text{\AA}$. The K and K$'$ points of WSe$_2$ were folded onto the K and K$'$ points of the heterostructure.
Exchange and correlation effects were described within the generalized gradient approximation using the Perdew-Burke-Ernzerhof functional\cite{PhysRevLett.77.3865}. A plane wave energy cutoff of 350 eV was used. Spin-polarized DFT+$U$ calculations were performed with a Hubbard \textit{U} parameter of 1.8 eV\cite{PhysRevB.57.1505}, including spin-orbit coupling. A ferromagnetic spin configuration was used to model the field-aligned Nb$_3$Cl$_8$ moments relevant to the high-field valley polarization measurements. A k-point mesh of $9\times9\times1$ was used for all calculations. 
%The pyband package was used to plot the projected-bands.\\

\bigskip
\noindent\textbf{Data availability}\\
\noindent
All relevant data are available in the main text, Supplementary Information, or upon request to the authors.

\bigskip
\noindent\textbf{Acknowledgment}\\
\noindent
This work was supported by the National Natural Science Foundation of China (No. 12425402) and the National Key R\&D Program of China (No. 2025YFA1411002 and No. 2022YFA1203902). H.C. and W.Y. acknowledge support from National Natural Science Foundation of China (No. 12425406), Research Grant Council of Hong Kong (HKU SRFS21227S05), and New Cornerstone Science Foundation. K.W. and T.T. acknowledge support from the JSPSKAKENHI (No. 21H05233 and No. 23H02052) and World Premier International Research Center Initiative (WPI), MEXT, Japan.  \\

\bigskip
\noindent\textbf{Author contributions}\\
\noindent
Y.Y. and X.H. conceived the project and designed the experiments. X.H., X.T., and Y.H. prepared the samples and fabricated the WSe$_2$/Nb$_3$Cl$_8$ devices. X.H. and X.T. performed the optical measurements, supported by Y.Z. and Y.G.. H.C. and C.X. carried out the first-principles calculations under the supervision of W.Y.. Z.C. grew the WSe$_2$ single crystals. Z.M. synthesized the Nb$_3$Cl$_8$ bulk crystals under the supervision of Y.S.. K.W. and T.T. grew the \textit{h}-BN single crystals. X.H., X.T., and H.C analyzed the results under the supervision of Y.Y and W.Y.. X.H. and Y.Y. wrote the manuscript. All authors discussed the results and contributed to the manuscript. \\

\bigskip
\noindent\textbf{Competing interests}\\
\noindent
The authors declare no competing interests.\\

\bigskip
\noindent\textbf{Additional information}\\
\noindent
\textbf{Supplementary information} The online version contains Supplementary material available at URL.\\

\normalem
\bibliographystyle{naturemag}
\bibliography{ref}

\end{document}